\documentclass[a4paper,11pt]{article}
\usepackage{jheppub} 
\usepackage{lineno}

\usepackage{graphicx}%
\usepackage{multirow}%
\usepackage{amsmath,amssymb,amsfonts}%
\usepackage{amsthm}%
\usepackage{mathrsfs}%
\usepackage[title]{appendix}%
\usepackage{xcolor}%
\usepackage{textcomp}%
\usepackage{manyfoot}%
\usepackage{booktabs}%
\usepackage{algorithm}%
\usepackage{algorithmicx}%
\usepackage{algpseudocode}%
\usepackage{listings}%

\usepackage{bbold}
\usepackage{amsmath}
\usepackage{amssymb}
\usepackage{comment}
\usepackage{graphicx}

\usepackage{tikz-feynman}
\usepackage{subcaption}
\tikzfeynmanset{compat=1.0.0}
\usepackage{comment}
\usepackage{wasysym}
\usepackage{mathrsfs}
\usepackage{slashed}
\usepackage{braket}
\usepackage[makeroom]{cancel}
\usepackage{amsthm}
\usepackage{wasysym}
\usepackage{mathrsfs}
\usepackage[makeroom]{cancel}
\usepackage{bbm}
\usepackage{bbold}

\DeclareMathOperator{\Tr}{Tr}
\newcommand{\hcal}{\mathscr{H}}
\newcommand{\Reals}{\mathbbm{R}}
\newcommand{\Complexes}{\mathbb{C}}

\newcommand{\lag}{\mathcal{L}}
\newcommand{\G}{\mathcal{G}}
\newcommand{\h}{\mathcal{H}}
\newcommand{\Dmu}{\partial_{\mu}}

\newcommand{\Dmuu}{\partial^{\mu}}

\newcommand{\TSigma}{\Tilde{\Sigma}}
\newcommand{\boldH}{\boldsymbol{H}}
\newcommand{\pib}{\boldsymbol{\pi}}

\title{Higgs radial modes and the UV completion of non-linear sigma models}

\author[a,b,c]{Giulio Marino}
\affiliation[a]{Dipartimento di Fisica E. Fermi, Università di Pisa, Largo B. Pontecorvo 3, I-56127 Pisa, Italy}
\affiliation[b]{INFN, Sezione di Pisa, Largo Bruno Pontecorvo 3, I-56127 Pisa, Italy}
\affiliation[c]{Dipartimento di Fisica, Sapienza Università di Roma, Italy}
\emailAdd{giulio.marino@phd.unipi.it}

\abstract{
We investigated non-linear sigma models with cosets $\G/\h$ represented by $SU(N)\times SU(N)/SU(N)$, $SU(N)/SO(N)$, and $SU(2N)/USp(2N)$. These models exhibit a transition to a strongly coupled regime above a threshold energy scale $\Lambda^*$, where the effective field theory approach breaks down. To address this issue, we introduced the so called Higgs radial modes within the Linear Sigma Model, crucial for restoring perturbative unitarity. Analyzing the $N>2$ case, we first notice that all radial modes coming from the most general irreducible representation $\hcal$ are necessary for unitarizing scattering amplitudes and ensuring the weakly coupled nature of the theory. 
The phenomenology behind these extra radial modes recalls the physics of the Higgs in the electroweak sector of the Standard Model. In the case of a model describing electroweak symmetry breaking with a more complex pattern, despite strict constraints from electroweak precision tests, these modes could stabilize the Higgs potential and be a possible solution for the vacuum instability issue. 
}

\makeatletter
\gdef\@fpheader{}
\makeatother

\begin{document}
\maketitle
\flushbottom

\section{Introduction}\label{sec1}
Non-linear sigma models (NL$\Sigma$M) play a fundamental role in the content of particle physics as they are closely associated with the perturbative unitarity problem in the scattering amplitudes of Goldstone modes. In the context of our discussion, a non-linear sigma model is defined as a map from Minkowski spacetime to a non-linear target space, specifically we will concentrate on coset target spaces of the form $\G/\h$, where $\G$ represents the symmetry group of the Lagrangian while $\h$ the stability subgroup preserved by the vacuum. Therefore, at a given point in spacetime, the chart $\chi^a(x)$ furnishes a coordinate system across the non-linear target space. These fields correspond to the Nambu-Goldstone bosons (NGBs) linked to the symmetry breaking pattern $\mathcal{G}/\h$. In the lowest order of perturbation theory, the kinetic Lagrangian can be expressed as:
\begin{equation}
    \lag=g_{ab}(\chi)\Dmu\chi^a\Dmuu\chi^b+\dots
\end{equation}
To construct an Effective Field Theory (EFT) describing low-energy dynamics of the NGBs, the CCWZ formalism is employed \cite{CCWZ1}\cite{CCWZ2}. However, above a threshold energy scale $\Lambda^*$, such theories become strongly coupled, since the perturbative expansion in powers of derivative becomes out of control. This is a symptom of the non-renormalizability of the NL$\Sigma$Ms and the corresponding energy growing behaviour of the scattering amplitudes involving the NGBs.

 The L$\Sigma$M is therefore defined as a renormalizable theory where new degrees of freedom are introduced by hand (with tuned coupling values) in order to address the perturbative unitarity issue of the scattering amplitudes, to provide a linear irreps of $\G$ and to make the theory weakly coupled\footnote{We want to emphasize that when discussing the concept of "UV completion", we are considering the possibility that the L$\Sigma$M theory extends the applicability of the low-energy EFT beyond the threshold energy scale $\Lambda^*$. However, it's important to highlight that the renormalization group evolution of the coupling might result in the theory becoming strongly coupled once more, such as in the case where a Landau pole emerges.}. 
 
 In this paper we will analyze three different cosets: 
 \begin{equation}
 SU(N)\times SU(N)/SU(N),\quad SU(N)/SO(N),\quad SU(2N)/USp(2N).    
 \label{eq:SSBpatterns}
 \end{equation}
 It is well known that if the symmetry breaking pattern is $SO(N)/SO(N-1)$, then it is sufficient to add just a singlet radial mode to construct a renormalizable and weakly coupled theory. This is the case of $SU(2)\times SU(2)/SU(2)$, $SU(2)/SO(2)$ and $SU(4)/USp(4)$ where the patterns can be mapped one to one to $SO(4)/SO(3)$, $SO(3)/SO(2)$ and $SO(6)/SO(5)$ respectively. Indeed in the $SO(N)/SO(N-1)$ case, since the tensorial representations of $SO(N)$ are real, we can always impose real constraints in order to lower the number of physical degrees of freedom.
We will generalize the mechanism for the $N>2$ case and we will describe how all the radial modes have to be considered to cure the energy growing behaviour of the scattering amplitudes.

\subsection{Model independent coset construction}
In this section we will perform a model independent coset construction, whose validity is true for all the three patterns in \ref{eq:SSBpatterns}. Let us consider a generic spontaneous symmetry breaking (SSB) pattern $\G\rightarrow\h$ and denote with $G_\alpha$ the generators of the group $\G$ ($\alpha=1,\dots,n_G)$ with $T_i$ the generators of the unbroken group ($i=1,\dots,n_H)$ and with $X_a$ the generators belonging to the algebra of the coset $\G/\h$ ($a=1,\dots,n_G-n_H$). We will focus on the scenario where the quotient space $\G/\h$ exhibits symmetry, meaning there exists an automorphism $\mathcal{R}$ of the algebra (grading) under which the broken generators flip sign and the structure constants $f^{abc}$ vanish:
\begin{equation}
    \exists\mathcal{R}:\quad g\rightarrow\mathcal{R}(g)\quad |\quad T_i\rightarrow T_i,\quad X_a\rightarrow-X_a.
\end{equation}
Given the coset symmetry, the transformation rule induced on the NGB fields $\chi^a$ is
\begin{equation}
    \mathcal{P}_{L/R}:\quad \chi^a\rightarrow-\chi^a,
\end{equation}
which is an accidental symmetry of the NL$\Sigma$M Lagrangian broken by the Wess-Zumino-Witten (WZW) term at higher orders. Particularly this implies that, at the lowest order, any process involving an odd number of NGBs will vanish.

From Goldstone theorem \cite{Nambu1}\cite{goldstone}, we expect $n_G-n_H$ NGBs which can be parameterized by introducing a complex matrix valued field denoted as $\Sigma$, defined as:
\begin{equation}
    \Sigma=\exp\bigg(2i\frac{\chi^aX^a}{f}\bigg).
\end{equation}
Following the CCWZ prescription, we can write down the most general Lagrangian at the lowest order in perturbation theory:
\begin{equation}
    \lag^{(2)}=\frac{1}{2}(\Dmu\chi^a)^2-\frac{1}{6}\chi^a\Dmu\chi^b\chi^c\Dmu\chi^d f^{abi}f^{cdi},
    \label{eq:NLSMLag}
\end{equation}
where the $f$ structure constants are specified by the structure of the group under investigation. However, since we are interested in the patterns in \eqref{eq:SSBpatterns} and since all of them involve constant structures whose indices runs over $SU(N)$, we can manipulate the $f^{abi}f^{cdi}$ term using an useful relation for the $SU(N)$ algebra \cite{Haber}
\begin{equation}
    f^{abe}f^{cde}=\frac{2}{N}(\delta_{ac}\delta_{bd}-\delta_{ad}\delta_{bc})+d_{ace}d_{bde}-d_{bce}d_{ade}.
    \label{eq:ffrelation}
\end{equation}
and since the symmetric d-symbols can be rewritten in terms of invariant tensors of $SU(N)$
\begin{equation}
    d^{abe}d^{cde}=A\delta^{ab}\delta^{cd}+B(\delta^{ac}\delta^{bd}+\delta^{ad}\delta^{bc}).
    \label{eq:dddecomp}
\end{equation}
the relation \eqref{eq:ffrelation} becomes
\begin{equation}
    f^{abi}f^{cdi}=\Tilde{C}(\delta^{ac}\delta^{bd}-\delta^{ad}\delta^{bc})
\end{equation}
where 
\begin{equation}
    \Tilde{C}=A-B+\frac{2}{N}.
\end{equation}
It is important to note that the expansion $f^{abi}f^{cdi}$ is applicable to all patterns stemming from $\G=SU(N)$, with distinct values for $A$ and $B$ determined by the specific coset structure. With this, we can proceed to calculate the two-to-two scattering amplitude $\mathcal{A}(\pi^a\pi^b\rightarrow\pi^c\pi^d)$, deriving a generale expression applicable to all three patterns under consideration. Bose and crossing symmetry imply that the scattering amplitude given by the Lagrangian \eqref{eq:NLSMLag} respects the general decomposition:
\begin{equation}
    \mathcal{A}(\pi^a\pi^b\rightarrow\pi^c\pi^d)=\Bar{A}(s,t,u)\delta^{ab}\delta^{cd}+\Bar{A}(t,s,u)\delta^{ac}\delta^{bd}+\Bar{A}(u,t,s)\delta^{ad}\delta^{bc}+\Bar{B}(s,t,u)\epsilon^{abcd}
    \label{eq:AmpliDec}
\end{equation}
where s,t,u are the standard Mandelstam variables
\[
s=(p_a+p_b)^2,\quad t=(p_a-p_c)^2,\quad u=(p_a-p_d)^2.
\]
\begin{figure}[h!!!!!!!!]
    \centering
    \includegraphics[scale=1.5]{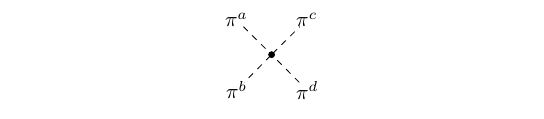}
    \caption*{}
    \label{fig:enter-label}
\end{figure}

Furthermore, crossing symmetry necessitates that $\Bar{B}$ be antisymmetric when exchanging two Mandelstam variables. However, at this stage, we disregard this contribution, as it only arises when $N=4$. In conclusion one arrives to the non-linear sigma model $\pi\pi$ scattering amplitude\footnote{Notice that we are considering the massless limit of the theory, $\sum_{\alpha}p_\alpha^2=0$ in the on-shell amplitude.}:
\begin{equation}
\Bar{A}(s,t,u)=i\frac{\Tilde{C}}{f^2}s+\dots
\end{equation}

\section{$\mathbf{SU(N)\times SU(N)/SU(N)}$ Sigma Model}\label{sec2}
In this section we focus on the QCD-like case and we analyze the mechanism through which the Linear Sigma Model solves the perturbative unitarity issue arising from the scattering amplitudes. At this purpose new radial modes are added in the theory and we will prove that all of them have to be considered to make the theory weakly coupled.
\subsection{From Non Linear to Linear Sigma Model}
Let us start constructing the NL$\Sigma$M for the more general $N$ flavour chiral pattern:
\begin{equation}
    SU(N)_L\times SU(N)_R \times U(1) \rightarrow SU(N)_V
\end{equation}
where the extra abelian symmetry is added in order to help us in recognizing the physical nature of the degrees of freedom.
From Goldstone theorem  we expect $(N^2-1)+1$ NGBs and they can be parameterized as 
\begin{equation}
\Tilde{\Sigma}=\Sigma \exp\bigg(i\frac{\eta}{f}\bigg)=\exp\bigg(2i\frac{\pib}{f}\bigg)\exp\bigg(i\frac{\eta}{f}\bigg),
\end{equation}
where $\pib=\pi^{\hat{a}}X^{\hat{a}}$ and $\eta$ transform respectively as the adjoint representation and a singlet of unbroken subgroup $\h$.
Furthermore, it is possible to introduce the transformation rule of $\TSigma$ under the group $\G$ as:
\begin{equation}
    \TSigma(x)\rightarrow UL\TSigma(x)R^\dagger,\quad U\in U(1), \quad L\in SU(N)_L,\quad R\in SU(N)_R. 
\label{eq:chTransf}
\end{equation}
As expected, the action of $\G$ reduces to a linear transformation when we restrict to the stability subgroup $\h$. By construction, $\TSigma$ has to satisfy the following $N^2$ non-linear real constraints
\begin{equation}
    \TSigma^\dagger\TSigma=\TSigma\TSigma^\dagger=\mathbbm{1}.
    \label{eq:nonlinearconstraintSUN}
\end{equation}
At this stage, we can provide a naive counting of the physical degrees of freedom. We started from $\TSigma\sim\text{Mat}(N,\Complexes)$, which by definition describes $2N^2$ real degrees of freedom, but $N^2$ of them are removed from the non-linear conditions (\ref{eq:nonlinearconstraintSUN}). Therefore, we conclude that $N^2$ of them remain and can be arranged according to irreducible representations of the unbroken group:
\begin{equation}
    N^2_\Reals=\underbrace{N^2-1}_{\pi^{\hat{a}}(x)}+\underbrace{1}_{\eta(x)}.
\end{equation}
As we expand the lowest-order Lagrangian, we observe that the $\eta$ field appears only in the kinetic term. Therefore, if we perform a field redefinition to standardize the normalization factor for the kinetic term, the NL$\Sigma$M Lagrangian can be expressed as shown in equation \eqref{eq:NLSMLag}.

Now, our goal is to make the theory weakly coupled by introducing additional degrees of freedom. Thus, we start considering a linear representation $\hcal$ of $\G$ in the linear space of $N\times N$ complex matrices.
For $N>2$ one may think that, since the fundamental and anti-fundamental representations of $SU(N)$ are not linked by a pseudo-reality condition, no linear constraints intervene in lowering the number of physical degrees of freedom. Hence, one might consider that all the additional radial modes are necessary to reinstate perturbative unitarity, given that the smallest linear representation contains more then one radial mode. To prove this assertion, we opt to employ chiral perturbation theory, expanding the Linear Sigma Model Lagrangian at the lowest order and checking the energy behavior of the relevant scattering amplitudes.
As already said, we start by considering a linear representation $\hcal$ in Mat($N,\Complexes$) which describes $2N^2_\Reals$ degrees of freedom and organizes them in terms of irreps of the unbroken group $SU(N)_V$:
\[
2N^2_\Reals = \underbrace{N^2-1}_{\pi^a(x)}+\underbrace{1}_{\eta(x)}+\underbrace{N^2-1}_{H^a(x)}+\underbrace{1}_{h(x)}
\]
where $h(x)$ is the Higgs singlet scalar radial mode already present in the two flavour case and $H^a(x)$ are the new Higgs fields that live in the adjoint representation. We recall that $\pi^a(x)$ are the $N^2-1$ NGBs arising from the chiral part of the symmetry breaking and $\eta(x)$ is the Goldstone associated to the abelian symmetry breaking.

Using polar decomposition, which tells us that a generic matrix in Mat($N,\Complexes$) can be decomposed as the product of an hermitian $\phi$ matrix times a unitary matrix $\TSigma$, we parameterize $\mathscr{H}$ as:
\begin{equation}
    \mathscr{H}=\TSigma\phi,\quad \TSigma^\dagger\TSigma=\mathbbm{1},\quad \phi^\dagger=\phi.
\end{equation}
So we can finally write\footnote{This parametrization is arbitrary and we could have chosen another one, in such a way that the expansion around the origin is the same.}:
\begin{equation}
    \mathscr{H}=\Tilde{\Sigma}(x)\phi(x)=\TSigma\cdot\phi_H\phi_h=e^{\frac{2i}{f}\pi^aX^a}e^{\frac{i}{f}\eta}\bigg(1+\frac{H^a}{f}X^a\bigg)\bigg(f+h\bigg)
    \label{eq:hcalpara1}
\end{equation}
which allows us to consider the most general Linear Sigma Model Lagrangian as:
\begin{equation}
\mathcal{L}=\underbrace{\frac{1}{4}\Tr[\Dmu\mathscr{H}^\dagger\Dmuu\mathscr{H}]}_{\{1\}}-\underbrace{V(\mathscr{H}\mathscr{H}^\dagger)}_{\{2\}}+\underbrace{\text{Pol}(\det\mathscr{H})}_{\{3\}}    
\label{eq:lsmLag}
\end{equation}
which is manifestly renormalizable, if we neglect the last term. 
Moreover we notice that the field $\mathscr{H}$ transforms under $\mathcal{G}$ as:
\[
\mathscr{H}\xrightarrow{SU(N)_L\times SU(N)_R}L\mathscr{H}R^\dagger,\quad \mathscr{H}\xrightarrow{U(1)}e^{i\alpha}\mathscr{H}
\]
where $L\in SU(N)_L$ and $R\in SU(N)_R$ and we choose to orient the vacuum expectation value as
\[
\langle\mathscr{H}\rangle=f\mathbbm{1}
\]
in order to perform the breaking in the right way.
One can notice that $\mathcal{L}$ without \{3\} is invariant under the global symmetry group $\G$, but if we introduce it, then only chiral symmetry is preserved and the $U(1)$ abelian symmetry is explicitly broken. Since the determinant is already a polynomial of grade $D=N$. Notice that \{3\} is non-renormalizable for $N>4$; in that case the abelian $U(1)$ symmetry is accidental. If present in the Lagrangian, the explicitly breaking of $U(1)$ lead to a mass contribution for the $\eta$ field.

We want to stress that, the reason why we are considering an extra abelian symmetry in $\G$ is just to associate the particle $\eta$ to a NGB arising from a symmetry breaking, however if we had considered the pure $N$-flavour chiral symmetry breaking pattern, it would have been associated to a massive pNGB arising from the explicitly breaking given by \{3\}.

Moreover we notice that since $\TSigma$ and $\phi$ do not commute then \eqref{eq:lsmLag} is no more symmetric under $\mathcal{P}_{L/R}$ transformation on the pion field. One may verify that if we act with the automorphism $\mathcal{R}$ we have that
\begin{equation}
    \mathscr{H}\rightarrow\mathscr{H}'=\TSigma^\dagger\bigg(1-\frac{H^a}{f}T^a\bigg)\bigg(f+h\bigg)\neq\hcal^\dagger.
\end{equation}
Thus, $\mathcal{P}_{L/R}$ is accidental only in the NL$\Sigma$M. This means that, when we will perform the chiral perturbative expansion, there may be terms with an odd numbers of $\pi$ and $\eta$ fields, even though in the NL$\Sigma$M this was not allowed (at the lowest order).

We revisit the parametrization for $\hcal$ as defined in equation \eqref{eq:hcalpara1} and proceed to expand the three distinct terms in \eqref{eq:lsmLag}. Additionally, we perform a field redefinition to normalize canonically the kinetic terms. Consequently, the L$\Sigma$M Lagrangian takes the following form:
\begin{equation}
    \mathcal{L}=\mathcal{L}_{kin}+\mathcal{L}_{\pi\pi}+\mathcal{L}_{h\pi}+\mathcal{L}_{H\pi}+\mathcal{L}_{hH}+\mathcal{L}_\eta+\mathcal{L}_{\text{mix}}.
\end{equation}
where the explicit form can be found in Appendix \ref{app:A}.

\subsection{\label{sec:AdjHiggses}The role of adjoint Higgs bosons in the UV completion of the NL$\Sigma$M}
In the preceding section, we have derived an explicit formulation for the Lagrangian of the $SU(N)\times SU(N)/SU(N)$ Linear Sigma Model. In this section we compute explicitly the $\pi\pi$ scattering amplitude and we will analyze the way in which the adjoint Higgs radial modes intervene in making the theory weakly coupled.
The main difference to the NL$\Sigma$M is the new interaction term involving the extra radial modes $H^a$:
\begin{equation}
    \mathcal{L}_{h\pi^i\pi^j}=\sqrt{\frac{2}{N}}\frac{1}{f}\Dmu\pi^i\Dmuu\pi^jh\delta^{ij},
\end{equation}
\begin{equation}
    \mathcal{L}_{\pi^i\pi^jH^k}=\frac{1}{f}\Dmu\pi^i\Dmuu\pi^jH^kd^{ijk}.
\end{equation}
Analyzing the total $\pi\pi$ scattering amplitude, one finds that there is a contribution coming from the NL$\Sigma$M, $i.e.$ the four pion vertex interaction, and new contributions where the radial modes mediate the $\pi\pi$ scattering
\begin{equation}
    \mathcal{A}_{tot}=\mathcal{A}_\pi+\mathcal{A}_h+\mathcal{A}_H.
\end{equation}

\begin{figure}[h!!!!!!!!]
    \centering
    \includegraphics[scale=1.5]{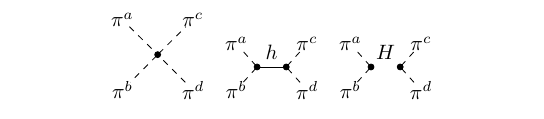}
    \caption*{}
    \label{fig:enter-label}
\end{figure}
After some manipulations, in the massless limit, we arrive to the expression for  the total $\Bar{A}(s,t,u)$ 
\begin{equation}
    \Bar{A}_{\text{tot}}(s,t,u)=\frac{i}{f^2}\bigg(\Tilde{C}-\frac{2}{N}-(A-B)\bigg)s-\frac{i}{f^2}\bigg((A+2B)m^2_H-\frac{2}{N}m^2_h\bigg)+\mathcal{O}\bigg(\frac{m^4_H}{s},\frac{m^4_h}{s}\bigg).
    \label{eq:Abar}
\end{equation}
where $\Tilde{C}$ is defined as in \eqref{eq:Ctilde} and $A$, $B$ are the coefficients defined in the previous section. The first easy check could be done restricting down to the $N=2$ case, where the $H$ dependence disappear because of $A+2B=0$ from \eqref{eq:A-B,A+2B}. At this point we can finally arrive to the main result
\begin{equation}
\Tilde{C}-\frac{2}{N}-(A-B)=0    
\end{equation}
which means that the perturbative unitarity issue and the growing energy behaviour of the total amplitude are cured by taking into account all the radial modes provided by $\hcal$, both the singlet $h(x)$ and those living in the adjoint representation $H(x)$. Given that even a single scattering amplitude requires the presence of all radial modes, and considering that the representation for $\hcal$ is as generic as possible, we conclude that in order to restore perturbative unitarity and render the theory weakly coupled, all radial modes are necessary.

\begin{equation}
    \mathcal{A}^{\text{tot}}(\pi^a\pi^b\rightarrow\pi^c\pi^d)=\mathcal{A}_\pi+\mathcal{A}_h+\mathcal{A}_{H}\sim\text{const}+\mathcal{O}\bigg(\frac{m^4_H}{s},\frac{m^4_h}{s}\bigg).
\end{equation}

An additional consistency check can be performed by observing that all other scattering amplitudes exhibit a constant energy behavior at the lowest order. Table \ref{tab:my_label} provides a list of some potential scattering processes allowed by the L$\Sigma$M interactions.
\begin{figure}[h!!!!!!!!]
    \centering
    \includegraphics[scale=1.5]{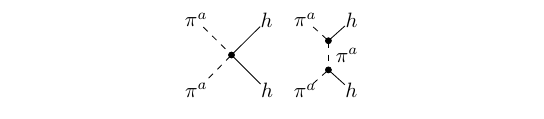}
    \caption*{}
    \label{fig:enter-label}
\end{figure}

\begin{table}[h!]
    \centering
    \begin{tabular}{c c}
        $\pi\pi\rightarrow\eta\eta$&$\pi\pi\rightarrow HH$\\
        $\pi\pi\rightarrow\eta \pi$&$\pi\pi\rightarrow hH$\\
        $\pi h\rightarrow \pi H$&$\pi h\rightarrow \eta H$\\
        $\pi\pi\rightarrow hh$&$\pi h\rightarrow \pi h$\\
        $\pi H\rightarrow \pi H$&$\pi H\rightarrow \eta H$\\
        $\pi H\rightarrow \pi h$&$\pi H\rightarrow \eta h$\\
        $\eta \eta \rightarrow \eta \eta$&$\eta \eta\rightarrow hh$\\
        $\eta H\rightarrow \eta H$&$\eta h\rightarrow \eta h$\\
        $HH\rightarrow\pi\eta$&$\pi \pi\rightarrow\pi\pi$
    \end{tabular}
    \caption{Some possible scattering processes within the L$\Sigma$M.}
    \label{tab:my_label}
\end{table}
However, here we analyze also the $\pi\pi\rightarrow hh$ process. We recall the interactions:
\begin{equation}
    \begin{split}
        \lag_{h\pi^i\pi^j}&=\sqrt{\frac{2}{N}}\frac{1}{f}\Dmu\pi^i\Dmuu\pi^j h\delta^{ij},\\
        \lag_{\pi\pi hh}&=\frac{1}{Nf^2}(\Dmu\pi^a)^2h^2.
    \end{split}
\end{equation}
where the relevant Feynman diagrams for the process are summarized below.

In conclusion we arrive at 
\begin{equation}
    \mathcal{A}(\pi\pi\rightarrow hh)= -\frac{2i}{Nf^2}(s+t+u)\sim \text{const},
\end{equation}
which, as predicted, goes as a constant in energy. Thus, we observe that the $\pi\pi\rightarrow hh$ scattering is cured without keep in consideration the extra adjoint radial modes $H^a(x)$. Therefore, while not all scattering amplitudes are unitarized by considering all modes, our proof demonstrates that at least one process requires the inclusion of all Higgses. Consequently, we conclude that all radial modes provided by $\hcal$ are indispensable.
One has to repeat the whole story also for the other processes as the $\pi\pi\rightarrow HH$ scattering where we write only the relevant Feynman diagrams.

We emphasize that interactions stemming from the perturbative expansion of the potential have been disregarded in the calculation of the scattering processes, as they do not contribute to an energy-dependent growth in the amplitudes.

\subsection{\label{sec:Potential and mass terms}Potential and mass terms}
In this section we will focus on the potential term of the Lagrangian density \eqref{eq:lsmLag}. It is particularly interesting because it provides information about the masses of the radial modes. 
The most general potential that one can write\footnote{In the following we choose to neglect the $\det(\hcal)$ term in the potential.} has the form:
\begin{equation}
    V(\mathscr{\hcal^\dagger\hcal})=-\frac{\mu^2}{4}\Tr(\hcal^\dagger\hcal)+\frac{1}{16}\bigg(\lambda_1[\Tr(\hcal^\dagger\hcal)]^2+\lambda_2\Tr(\hcal^\dagger\hcal\hcal^\dagger\hcal)\bigg)
\end{equation}
where $\mu,\lambda_1$ and $\lambda_2$ are three real coefficients.
For instance, let us start proving the orientation of the vacuum expectation as
\begin{equation}
\langle\hcal\rangle=f\mathbbm{1}.
\label{eq:vev}
\end{equation}
Minimizing the potential, we get the stable configurations:
\begin{equation}
\frac{\partial V}{\partial (\hcal^\dagger\hcal)_{ij}}=-\frac{\mu^2}{4}\delta_{ij}+\frac{1}{16}\bigg[2\lambda_1\delta_{ij}\Tr(\mathcal{\hcal^\dagger\hcal})+2\lambda_2(\hcal^\dagger\hcal)_{ij}\bigg]\overset{!}{=}0.
\end{equation}
Since $\hcal^\dagger\hcal$ is Hermitian we can always diagonalize it by an unitary transformation, therefore its eigenvalues will be of the form 
\begin{equation}
    h_i^\dagger h_i=f^2_i.
\end{equation}
Thus, the first derivative of the potential becomes:
\begin{equation}
    \frac{\partial V}{\partial (\hcal^\dagger\hcal)_{ij}}=\bigg[-\frac{\mu^2}{4}+\frac{\lambda_1}{8}\bigg(\sum_if^2_i\bigg)\bigg]\delta_{ij}+\frac{\lambda_2}{8}\text{diag}(f^2_1,f^2_2,\dots,f^2_n)=0.
\end{equation}
At this point, we can solve the equation for each $f^2_i$ and we find an explicit expression for $f$:
\begin{equation}
    f_i\equiv f=\sqrt{\frac{2\mu^2}{(\lambda_1 N+\lambda_2)}}, \quad \forall f_i.
\end{equation}
Referring back to the field parametrization \eqref{eq:hcalpara1} we perform the perturbative expansion and after the usual field redefinition we can finally write the mass Lagrangian as:
\begin{equation}
    \mathcal{L}_{\text{mass}}=\frac{1}{2}m_H^2H^aH^a+\frac{1}{2}m_h^2h^2
\end{equation}
where
\begin{equation}
    m^2_h=2\mu^2,\quad m_H^2=2\mu^2\frac{\lambda_2}{\lambda_1N+\lambda_2}.
\end{equation}
Moreover one could derive also the interaction terms and the results will lead us to the following Feynman diagrams.

\begin{figure}[h!!!!!!!!]
    \centering
    \includegraphics[scale=1.5]{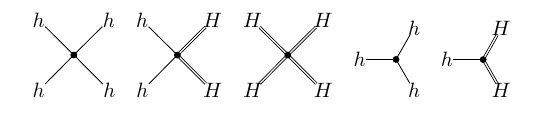}
    \caption*{}
    \label{fig:enter-label}
\end{figure}

We conclude that, for fixed values of $\mu,\lambda_1$ and  $\lambda_2$, the mass of the extra radial modes becomes smaller as the number of flavours increases. In the large $N_f$ limit, the adjoint radial modes become massless, unlike the singlet $h$, which remains massive. Investigating the behavior of the theory in this limit would be instriguing. Upon examining the Lagrangian \eqref{eq:allLag} we conclude that in the large $N_f$ limit, the only non zero terms are $\lag_{H\pi}$ and $\lag_{\pi\pi}$, resulting in complete decoupling of the singlet mode from the theory. Thus, the L$\Sigma$M theory splits as a massive scalar for $h(x)$ and two adjoint theories coupled each other.

\section{$\mathbf{SU(N)/SO(N)}$ and $\mathbf{SU(2N)/USp(2N)}$ Sigma Models}
\label{chap:5}
Previously we analyzed a QCD-like theory with a focus on NL$\Sigma$M and its perturbative unitarity issue. We showed that, in the more general case, new extra radial modes must be considered in order to make the theory weakly coupled and linearly realized at high energy scales. At this point we would like to repeat the whole story also for other patterns. For example, if we decide to choose a 2nd-rank symmetric tensor representation for the scalar field, the $SU(N)$ group can break into $SO(N)$, instead if we choose a 2nd-rank antisymmetric tensor representation it is possible to break $SU(2N)$ into the unitary symplectic group $USp(2N)$. In the following we will study the coset construction for both patterns and their completion through L$\Sigma$M. In section \ref{sec:SUN/SON Coset Construction} we will apply the coset construction procedure to the $SU(N)/SO(N)$ symmetry breaking pattern. In section $\ref{sec:SU(2N)/USp(2N)}$ we will give a brief introduction to the $SU(2N)/USp(2N)$ sigma model, stressing the similar procedure with respect to the $SU(N)/SO(N)$ case.

\subsection{\label{sec:SUN/SON Coset Construction}$SU(N)/SO(N)$ Coset construction}

Let us start considering the CCWZ steps in order to construct a $SU(N)/SO(N)$ NL$\Sigma$M.
In the following we will denote with $T^i$ ($i=1,\dots,\frac{N(N-1)}{2}$) the unbroken generators belonging to the algebra of $\h$, and with $X^a$ ($a=1,\dots,\frac{N(N+1)}{2}-1$) the broken generators belonging to the algebra of the coset $\G/\h$. From the SSB we expect $\frac{N(N+1)}{2}-1$ NGBs that we can parameterize as a symmetric traceless matrix.
First we identify the automorphism of the algebra that makes the coset space symmetric. It's notable that upon defining $\mathcal{R}$ as
\begin{equation}
\mathcal{R}:\quad g\rightarrow\mathcal{R}(g)=(g^{-1})^T
\end{equation}
the algebra transforms under its action as
\begin{equation}
    X^a\rightarrow-X^a,\quad T^i\rightarrow T^i.
\end{equation}

Given the count of NGBs, the choice for $\Sigma$ falls upon the symmetric traceless matrix representation
\begin{equation}
    \Sigma=e^{\frac{2i}{f}\pi^aX^a}
\end{equation}
where $\G$ acts as
\begin{equation}
    \Sigma\xrightarrow{\G}M\Sigma\mathcal{R}(M)^\dagger=M\Sigma M^T.
\end{equation}
By construction $\Sigma$ satisfies the following non linear constraints
\begin{equation}
    \begin{split}
        \Sigma^\dagger\Sigma=\Sigma\Sigma^\dagger=\mathbbm{1}&\Rightarrow N^2 \,\,\,\text{constraints,}\\
        \det(\Sigma)=1 &\Rightarrow 1 \,\,\,\text{constraint} 
    \end{split}
\end{equation}
and also the symmetric condition
\begin{equation}
    \Sigma=\Sigma^T,
    \label{eq:Sigmatransf}
\end{equation}
which adds $\frac{N(N-1)}{2}$ new independent constraints.
The naive counting of the physical degrees of freedom for $\Sigma$ is
\[
2N_\Reals^2-(N^2+1)_\Reals-\frac{N(N-1)}{2}=\underbrace{\frac{N(N+1)}{2}-1}_{\# NGB}.
\]
Performing the perturbative expansion up to $\mathcal{O}(p^2)$ order,  the lowest order Lagrangian can be written as in \eqref{eq:NLSMLag}.
The next step involves achieving UV completion of the theory to address the perturbative unitarity concern. This is accomplished by introducing a new field, denoted as $\hcal$, which provides a linear representation for the group $\G$. The minimal choice for $\hcal$ representation is the symmetric $N\times N$ complex matrix, since $\Sigma$ satisfy the \eqref{eq:Sigmatransf} condition. This matrix framework encompasses $2N^2 - N(N-1)$ degrees of freedom.

At this point, we would like to provide an useful parameterization for $\hcal$, and using the Takagi factorization \cite{Takagi1}\cite{Takagi2} we are able to decompose a generic $N\times N$ symmetric complex matrix $\mathscr{H}$ as
\begin{equation}
    \hcal=UDU^T
\end{equation}
where U is an unitary matrix and D is a diagonal matrix with positive eigenvalues. Furthermore we can decompose $U$ as
\begin{equation}
   \underbrace{U}_{U(N)}=\underbrace{e^{i\eta}}_{U(1)}\underbrace{\Sigma e^{i\alpha^iT^i}}_{SU(N)}
\end{equation}
and without loss of generality we can parameterize the $\hcal$ field as
\begin{equation}
    \hcal=(f+h)e^{\frac{i}{f}\eta}\Sigma(1+\mathbf{H})\Sigma,\quad \mathbf{H}=\frac{H^a X^a}{f} .
    \label{eq:Hpar}
\end{equation}
Before proceeding further, let's conduct some preliminary checks:
\begin{itemize}
    \item From equation \eqref{eq:Hpar} it is evident that $\hcal$ is symmetric, since also $\Sigma$ and $H$ are symmetric: 
        \begin{equation}
            \hcal^T=\hcal.
        \end{equation}
    \item The transformation rule of $\hcal$ under the action of $\G$ is defined as:
    \begin{equation}
        \hcal\xrightarrow{\G}V\hcal V^T,\quad V\in SU(N),
        \label{eq:transfHcal}
    \end{equation}
    which follows from 
    \begin{equation}
        \Sigma\xrightarrow{\G}V\Sigma V^T;\quad H\xrightarrow{\G}VHV^T.
    \end{equation}
\end{itemize}

Now we can perform the perturbative expansion of the L$\Sigma$M Lagrangian
\begin{equation}
    \mathcal{L}=\frac{1}{4}\Tr(\Dmu\hcal^\dagger\Dmuu\hcal)-V(\hcal^\dagger\hcal).
\end{equation}
and defining
\begin{equation}
    \Phi=\Sigma(1+\mathbf{H})\Sigma.
\end{equation} 
After some manipulations we finally arrive to the Lagrangian:
\begin{equation}
\begin{split}
    \mathcal{L}&=\frac{1}{4}\bigg[(f+h)\Dmu h \Tr(\Phi\Dmu\Phi^\dagger)+(\Dmu h)^2\Tr(\Phi\Phi^\dagger)+\frac{i}{f}(f+h)^2\Dmu\eta \Tr(\Phi\Dmu\Phi^\dagger-\Dmu\Phi \Phi^\dagger)+\\
    &+\frac{1}{f^2}(f+h)^2(\Dmu \eta)^2\Tr(\Phi\Phi^\dagger)+(f+h)^2\Tr(\Dmu\Phi\Dmu\Phi^\dagger) \bigg]
    \end{split}
    \label{eq:SU5/SO5 lag}.
\end{equation}
All the necessary tools are provided for computing the scattering processes. Furthermore, it can be proved that, as in the previous case, perturbative unitarity is restored by considering all the additional radial modes provided by the irresp $\hcal$.

\subsection{\label{sec:SU(2N)/USp(2N)}A brief introduction to $SU(2N)/USp(2N)$ Sigma model}
The same procedure can be applied to the $SU(2N)/USp(2N)$ SSB pattern and the results are quite similar. 
In particular we identify with the symplectic group the set of special $2N$-complex matrices which leave unchanged the symplectic matrix $\Omega$:
\[
USp(2N):\bigg\{M\in Mat(2N,\Complexes),\quad M^T\Omega M=\Omega,\quad \Omega=
\begin{bmatrix}
    0&\mathbbm{1}_N\\
    -\mathbbm{1}_N&0
\end{bmatrix},\quad \det M=1\bigg\}.
\]
With this definition, the $USp(2n)$ group has dimension
\begin{equation}
    \text{dim}[USp(2N)]=\frac{N(N+1)}{2}-1.
\end{equation}
Therefore the $N^2-1$ generators \{$G^\alpha$\} of $SU(N)$ can be split into the unbroken generators \{$T^i$\} belonging to $USp(2N)$ ($i=1,\dots,\frac{N(N+1)}{2}-1$) and the broken generators \{$X^a$\} belonging to the coset space $\G/\h$ ($a=1,\dots,\frac{N(N-1)}{2}$).
From Goldstone theorem we expect $\frac{N(N-1)}{2}$ NGBs, corresponding to the number of broken generators. 
Thus NGBs live in the antisymmetric representation of $SU(N)$. 

In particular we can parameterize the $\Sigma$ field as
\begin{equation}
    \Sigma = e^{2i\pib/f}\Omega
\end{equation}
which is skew-symmetric by construction, $\Sigma^T=-\Sigma$. Therefore $\Sigma$ encompasses 
\begin{equation*}
    2N^2-(N^2+1)-\bigg(\frac{N(N+1)}{2}-1\bigg)
\end{equation*}
number of physical degrees of freedom.
An useful parameterization for $\hcal$ could be
\begin{equation}
    \hcal=(f+h)e^{i\eta/f}\Sigma \Omega(\Omega+\mathbf{H}\Omega)\Sigma \Omega
\end{equation}
which is indeed a skew-symmetric matrix and describes $N(N-1)+2$ degrees of freedom.
As usual, we can expand the Lagrangian in perturbation theory, and the outcomes will align with those of the other two patterns. It's essential to include all Higgs radial modes to achieve a linear representation of $\mathcal{G}$ and resolve the perturbative unitarity issue.

\section{Conclusions}
We focused on the scenario of a chiral pattern $SU(N)\times SU(N)/SU(N)$, considering an arbitrary number of flavors, and explored the $SU(N)/SO(N)$ and $SU(2N)/USp(2N)$ cosets. While one might expect additional radial modes to exist, our results demonstrate that all radial modes within the most general irreducible representation $\hcal$ are necessary to make the theory weakly coupled, addressing the perturbative unitarity issue. By expanding the linear sigma model lagrangian to first order in perturbation theory and explicitly computing the relevant scattering amplitudes, we verified that only including all radial modes originating from the most general irreducible representation $\hcal$ of the symmetry group $\G$ the energy growth behavior observed in non linear sigma model amplitudes is cured.

Additionally, we observed intriguing behavior in the masses of adjoint resonances when analyzing the potential within the $SU(N)\times SU(N)/SU(N)$ linear sigma model. As the number of flavors increases, the masses of these resonances diminish, becoming massless in the large $N_f$ flavor limit. Consequently, upon analyzing the lowest-order expanded lagrangian \eqref{eq:allLag}, we observe that the linear sigma model splits into a massive scalar theory and two interacting adjoint theories. Further investigation into this limit could offer insights for potential UV completions.

Moreover, we explored the $SU(N)/SO(N)$ and $SU(2N)/USp(2N)$ cosets, where the most general irreducible representation $\hcal$ can be found in the space of symmetric and skew-symmetric $N\times N$ complex matrices rispectively. Furthermore, we derived the explicit expression for $\hcal$, confirming that even within these cosets, the minimal number of additional radial modes required to restore perturbative unitarity corresponds to the one provided by the most general irreducible representation of $\G$.

Future investigations could delve into the phenomenology surrounding the analyzed extra radial modes, reminiscent of the physics observed in the Higgs sector of the Standard Model's electroweak theory. Focusing on the chiral pattern $SU(N)\times SU(N)/SU(N)$, potential implications for a more generalized electroweak theory could be considered. Beginning with a more intricate symmetry breaking pattern ($N>2$), constraints imposed by electroweak precision data \cite{ewpt} would tightly limit new sources of custodial symmetry breaking, impacting both the $f$ scale and the number of flavors $N$, influencing the choice of symmetry breaking pattern. However, under the assumption that these constraints can be met, the Higgs potential would undergo significant modifications due to loops involving the new adjoint scalar modes. Particularly, through an analysis of the RG flow of the couplings, one could reasonably hypothesize that the intervention of these new radial modes would stabilize the Higgs potential, addressing the vacuum instability issue observed in the Standard Model \cite{VacInst1,VacInst2,VacInst3}. Indeed, the inclusion of new scalars at loop level is expected to positively contribute to the beta function of the quartic coupling for the singlet mode, countering the negative contribution arising from top quark loops.

In summary, our research extends the reach of the Linear Sigma Model to encompass complex symmetry patterns, offering valuable guidance for constructing weakly coupled models in contexts like electroweak symmetry breaking or dark matter models.

\section*{Acknowledgments}
I would like to express my gratitude to Roberto Contino and Paolo Panci for their invaluable discussions and insightful contributions to this work.
\newpage
\appendix
\section{Linear Sigma Model Lagrangian for $SU(N)\times SU(N)/SU(N)$}
\label{app:A}
Expanding the L$\Sigma$M Lagrangian \eqref{eq:lsmLag} in perturbation theory one obtains 
\begin{equation}
\mathcal{L}=\mathcal{L}_{kin}+\mathcal{L}_{\pi\pi}+\mathcal{L}_{h\pi}+\mathcal{L}_{H\pi}+\mathcal{L}_{hH}+\mathcal{L}_\eta+\mathcal{L}_{\text{mix}}.
\end{equation}
where
\begin{align}
    \begin{split}
        \mathcal{L}_{kin}&=\frac{1}{2}(\Dmu\pi^a)^2+\frac{1}{2}(\Dmu\eta)^2+\frac{1}{2}(\Dmu h)^2+\frac{1}{2}(\Dmu H^a)^2\\
    \mathcal{L}_{\pi\pi}&=-\frac{1}{6f^2}\pi^a\Dmu\pi^b\pi^c\Dmu\pi^df^{abi}f^{cdi}\\
    \mathcal{L}_{h\pi}&=\frac{1}{f}\sqrt{\frac{2}{N}}h(\Dmu\pi^a)^2+\frac{1}{Nf^2}h^2(\Dmu\pi^a)^2\\
    \mathcal{L}_{H\pi}&=\frac{1}{f}H^a\Dmu\pi^b\Dmu\pi^cd^{abc}+\frac{1}{Nf^2}H^aH^b\Dmu\pi^c\Dmu\pi^d\delta_{ab}\delta_{cd}+\\
    &+\frac{1}{2f^2}H^aH^b\Dmu\pi^c\Dmu\pi^dd^{ab\gamma}d^{cd\gamma}-\frac{1}{f^2}H^a\Dmu H^b\pi^c\Dmu\pi^df^{abi}f^{cdi}\\
    \mathcal{L}_{hH}&=\frac{1}{f}\sqrt{\frac{2}{N}}h\Dmu H^a\Dmuu H^a+\frac{1}{Nf^2}h^2(\Dmu H^a)^2+\frac{1}{f}\sqrt{\frac{2}{N}}\Dmu hH^a\Dmuu H^a+\\
    &+\frac{2}{Nf^2}h\Dmu hH^a\Dmuu H^a+\frac{1}{Nf^2}(\Dmu h)^2H^aH^a\\
    \mathcal{L}_\eta&=\frac{1}{f}\sqrt{\frac{2}{N}}h(\Dmu\eta)^2+\frac{1}{Nf^2}h^2(\Dmu\eta)^2+\frac{1}{Nf^2}(\Dmu\eta)^2H^aH^a+\\
    &+\frac{1}{f}\sqrt{\frac{2}{N}}\Dmu\eta H^a\Dmu\pi^a+\frac{4}{Nf^2}\Dmu\eta hH^a\Dmu\pi^a+\frac{1}{2f^2}\sqrt{\frac{2}{N}}\Dmu\eta H^aH^b\Dmuu\pi^c d^{abc}\\
    \mathcal{L}_{\text{mix}}&=\frac{2}{f^2}\sqrt{\frac{2}{N}}hH^a\Dmu\pi^b\Dmuu \pi^cd^{abc}.
    \label{eq:allLag}
    \end{split}
\end{align}

\section{Some manipulations over $\mathbf{d}$-symbols}
\label{App:Manipulation}
In this Appendix, we will manipulate the $d^{ab\gamma}d^{cd\gamma}$ term in order to fix the $A$ and $B$ coefficients in \eqref{eq:dddecomp}. Notice that the $\gamma$ index runs over $SU(N)$, $SO(N)$ and $USp(2N)$ subgroups for the three different patterns \eqref{eq:SSBpatterns} under investigations. 

In the QCD-like case, the index $\gamma$ runs over an $SU(N)$ subgroup of $\G$ and we can use the following two constraints \cite{Haber}:
\begin{equation}
    \begin{split}
        d^{ab\gamma}d^{ad\gamma}&=\frac{N^2-4}{N}\delta^{bd}\\
        \sum_a d^{aa\gamma}&=0.
    \end{split}
\end{equation}
This lead us to solve the following set of two equations:
\begin{equation}
    \begin{cases}
        A+BN^2&=\frac{N^2-4}{N}\\
        A(N^2-1)+2B&=0
    \end{cases}
\end{equation}
and the solution lead to the following result:
\begin{equation}
    A=-\frac{8-2N^2}{2N+N^3-N^5},\quad B=-\frac{4-5N^2+N^4}{2N+N^3-N^5}.
    \label{eq:ABcoeff}
\end{equation}
For future purpose we write down the following two relations:
\begin{equation}
    A-B=-\frac{2}{N}+\frac{N}{N^2-2},\quad A+2B=\frac{2(N^2-4)}{N(N^2+1)}
    \label{eq:A-B,A+2B}
\end{equation}
where both vanish if we restrict to the $N=2$ case. Thus $\Tilde{C}$ takes the following values:
\begin{equation}
    \Tilde{C}=\frac{N}{N^2-2}
    \label{eq:Ctilde}
\end{equation}
which reduce to $\Tilde{C}=1$ in the two flavour case, as it has to be.
 \bibliographystyle{JHEP}
 \bibliography{bibliography.bib}

\providecommand{\noopsort}[1]{}\providecommand{\singleletter}[1]{#1}%

\providecommand{\href}[2]{#2}\begingroup\raggedright\begin{thebibliography}{10}

\bibitem{CCWZ1}
S.R.~Coleman, J.~Wess and B.~Zumino, \emph{{Structure of phenomenological Lagrangians. 1.}}, \href{https://doi.org/10.1103/PhysRev.177.2239}{\emph{Phys. Rev.} {\bfseries 177} (1969) 2239}.

\bibitem{CCWZ2}
C.G.~Callan, Jr., S.R.~Coleman, J.~Wess and B.~Zumino, \emph{{Structure of phenomenological Lagrangians. 2.}}, \href{https://doi.org/10.1103/PhysRev.177.2247}{\emph{Phys. Rev.} {\bfseries 177} (1969) 2247}.

\bibitem{Nambu1}
Y.~Nambu, \emph{Axial vector current conservation in weak interactions}, \href{https://doi.org/10.1103/PhysRevLett.4.380}{\emph{Phys. Rev. Lett.} {\bfseries 4} (1960) 380}.

\bibitem{goldstone}
J.~Goldstone, \emph{Field theories with superconductor solutions}, {\emph{Il Nuovo Cimento (1955-1965)} {\bfseries 19} (1961) 154}.

\bibitem{Haber}
H.E.~Haber, \emph{{Useful relations among the generators in the defining and adjoint representations of SU(N)}}, \href{https://doi.org/10.21468/SciPostPhysLectNotes.21}{\emph{SciPost Phys. Lect. Notes} (2021) 21}.

\bibitem{Takagi1}
T.~Takagi{\emph{Japanese J. Math.} {\bfseries 1} (1927) }.

\bibitem{Takagi2}
R.A.~Horn and C.R.~Johnson, \emph{Matrix Analysis}, Cambridge University Press (1985), \href{https://doi.org/10.1017/CBO9780511810817}{10.1017/CBO9780511810817}.

\bibitem{ewpt}
R.o.P.P.~Particle Data Group~collaboration, \emph{{Review of Particle Physics}}, \href{https://doi.org/10.1093/ptep/ptaa104}{\emph{Progress of Theoretical and Experimental Physics} {\bfseries 2020} (2020) 083C01}.

\bibitem{VacInst1}
M.~{Sher}, \emph{{Electroweak Higgs potential and vacuum stability}}, .

\bibitem{VacInst2}
B.~Schrempp and M.~Wimmer, \emph{Top quark and higgs boson masses: Interplay between infrared and ultraviolet physics}, \href{https://doi.org/10.1016/0146-6410(96)00059-2}{\emph{Progress in Particle and Nuclear Physics} {\bfseries 37} (1996) 1–90}.

\bibitem{VacInst3}
U.~Ellwanger and M.~Lindner, \emph{Constraints on new physics from the higgs and top masses}, \href{https://doi.org/https://doi.org/10.1016/0370-2693(93)91164-I}{\emph{Physics Letters B} {\bfseries 301} (1993) 365}.

\end{thebibliography}\endgroup

\end{document}